\documentclass[journal]{IEEEtran}

\ifCLASSINFOpdf
\else
\fi

\usepackage{balance}
\usepackage[normalem]{ulem} 
\usepackage{amsmath,amsthm}
\usepackage{amssymb}
\usepackage{amsfonts}
\usepackage{comment}
\usepackage{lipsum}
\usepackage{mathtools}
\usepackage{cuted}
\usepackage[pdftex]{hyperref}

\usepackage{algorithmic}

\usepackage{array}
\usepackage{mdwmath}
\usepackage{mdwtab}
\usepackage{eqparbox}

\usepackage{url}
\usepackage{mathtools}
\usepackage{suffix}

\usepackage{ulem}
\usepackage{graphicx} 
\graphicspath{{../Figures/}} 

\usepackage{cuted}

\theoremstyle{definition}

\newtheorem{theorem}{Theorem}
\newtheorem{corollary}{Corollary}
\newtheorem{lemma}{Lemma}

\def\Pr{\mathrm{Pr}}
\def\Pout{P_{out}}
\def\Sb{S_\beta}

\makeatletter
\newsavebox\myboxA
\newsavebox\myboxB
\newlength\mylenA
\newcommand*\xoverline[2][0.9]{%
    \sbox{\myboxA}{$\m@th#2$}%
    \setbox\myboxB\null
    \ht\myboxB=\ht\myboxA%
    \dp\myboxB=\dp\myboxA%
    \wd\myboxB=#1\wd\myboxA
    \sbox\myboxB{$\m@th\overline{\copy\myboxB}$}
    \setlength\mylenA{\the\wd\myboxA}
    \addtolength\mylenA{-\the\wd\myboxB}%
    \ifdim\wd\myboxB<\wd\myboxA%
       \rlap{\hskip 0.5\mylenA\usebox\myboxB}{\usebox\myboxA}%
    \else
        \hskip -0.5\mylenA\rlap{\usebox\myboxA}{\hskip 0.5\mylenA\usebox\myboxB}%
    \fi}
\makeatother

\def\Pout{\textrm{P}_{out}}
\def\Poutasy{\textrm{P}_{out}^{asy}}
\def\Pe{{\overline{P_e}}}
\def\Peasy{{\overline{P_e}^{asy}}}
\def\Ce{C}
\def\Ceasy{C^{asy}}

\def\BraceTxt{L~\text{terms}}

\def\erfc{\textrm{erfc}}

 \DeclarePairedDelimiterX\MeijerM[3]{\lparen}{\rparen}%
{\begin{smallmatrix}#1 \\ #2\end{smallmatrix}\delimsize\vert\,#3}

\newcommand\Gfun[8][]{%
  G^{\,#2,#3}_{#4,#5}\MeijerM[#1]{#6}{#7}{#8}}

\WithSuffix\newcommand\Gfun*[7]{%
  G^{\,#1,#2}_{#3,#4}\MeijerM*{#5}{#6}{#7}}

\DeclarePairedDelimiterX\MeijerH[3]{\lparen}{\rparen}%
{\begin{smallmatrix}#1 \\ #2\end{smallmatrix}\delimsize\vert\,#3}

\newcommand\Hfun[8][]{%
  H^{\,#2,#3}_{#4,#5}\MeijerH[#1]{#6}{#7}{#8}}

\WithSuffix\newcommand\Hfun*[7]{%
  H^{\,#1,#2}_{#3,#4}\MeijerM*{#5}{#6}{#7}}

\usepackage{hyperref}
\hypersetup{pdftex,colorlinks=true,allcolors=blue}
\usepackage{hypcap}

\begin{document}
\title{\huge Selection Combining over Log-Logistic Fading Channels with Applications to Underwater Optical Wireless Communications}

\author{Yazan H. Al-Badarneh,~\IEEEmembership{Member,~IEEE}, Mustafa K. Alshawaqfeh,~\IEEEmembership{Member,~IEEE}, Osamah S. Badarneh,~\IEEEmembership{Member,~IEEE}
\thanks{Y. H. Al-Badarneh is with the Department of Electrical Engineering, The University of Jordan, Amman, 11942 (email: {yalbadarneh@ju.edu.jo)}}

\thanks{  M. K. Alshawaqfeh and O. S. Badarneh are with the Electrical Engineering Department, School of Electrical Engineering and Information Technology, German Jordanian University, Amman 11180, Jordan (e-mail: { mustafa.shawaqfeh, Osamah.Badarneh,}@gju.edu.jo).}}

\maketitle
\begin{abstract}
We study the performance of a selection combining (SC) receiver operating over independent but non-identically distributed log-logistic ($\mathcal{LL})$ fading channels. We first characterize the statistics of the output instantaneous signal-to-noise ratio (SNR) of the SC receiver. Based on the SNR statistics, we derive exact analytical expressions, in terms of multivariate Fox H-functions, for the outage probability, the average bit error rate, and the ergodic capacity. We also derive exact expressions for such performance measures when all channels are independent and identically distributed, as a special case.  Furthermore, we deduce simplified asymptotic expressions for these performance metrics assuming high values of average transmit SNR. To demonstrate the applicability of our theoretical analysis, we study the performance of an SC receiver in underwater optical wireless communication systems. Finally, we confirm the correctness of the derived analytical results using Monte Carlo Simulations.

\end{abstract}

\IEEEpeerreviewmaketitle

\section{Introduction}
Log-logistic ($\mathcal{LL})$ distribution has recently garnered much attention due to its effectiveness in fitting experimental data related to channel fluctuations in various communication systems. Several studies utilized the two-parameter $\mathcal{LL}$ distribution to accurately model different types of channel fluctuations, including turbulence fading in underwater optical wireless communication (UWOC) systems \cite{9580415}, misaligned gain in millimeter-wave cellular networks \cite{8628991}, and air-ground channels in unmanned aerial vehicle (UAV) communications \cite{8770066}. 

The performance of wireless communication systems in $\mathcal{LL}$ fading environment has been investigated very recently in \cite{electronics11152409}. The authors in \cite{electronics11152409} analyzed the outage probability and the ergodic capacity of a single-branch receiver.  However, a single-branch receiver is  highly prone to fading, which can significantly degrade the performance, especially in deep fading scenarios. Therefore, a multiple-branch receiver  can be employed to mitigate severe fading.

The performance of UWOC systems under different turbulence fading conditions was investigated in \cite{8606206}, \cite{8370053}. Recent advances in the UWOC systems have focused on the design and performance perspectives of Internet of underwater things (IoUT) systems \cite{9978925}, \cite{9826883}. However, analyzing the performance of UWOC systems with diversity reception is of practical importance. In this regard, employing a multiple-branch selection combining (SC) receiver in UWOC systems can combat turbulence fading and improve the performance. The SC receiver is characterized by its low complexity, as it selects the branch with the highest signal-to-noise ratio (SNR) among the available $L$ branches. In realistic communication settings, these $L$ branches may experience non-identical or identical fading statistics \cite{simon2005digital}.  

This letter mainly focuses on analyzing the performance of an $L$-branch SC receiver in $\mathcal{LL}$ fading environment with applications to UWOC systems. The key contributions of this work can be outlined as follows:

\begin{itemize}
\item We characterize the statistics of the output SNR of the $L$-branch SC receiver considering $L$ independent but non-identically distributed (i.n.i.d.) branches, as well as the special case of $L$ independent and identically distributed (i.i.d.) branches.

\item  We derive novel exact analytical expressions, in terms of multivariate Fox H-functions,  for the outage probability, the average bit error rate (BER), and the ergodic capacity. In addition, we obtain simple asymptotic expressions for such performance metrics assuming high values of average transmit SNR. 

\item We utilize the obtained expressions to study the performance of an UWOC system affected by temperature-induced turbulence. 

\end{itemize}

\section{  Statistics of the SC receiver output SNR }
We consider an $L$-branch SC receiver operating in $\mathcal{LL}$ fading environment. The output instantaneous SNR of the SC receiver is characterized by $ \gamma_{SC}=\max\{ \gamma_{1},  \gamma_{2}, . . . . , \gamma_{L} \}$, where $\gamma_{l}$ is  the instantaneous SNR of the ${l}$-th branch, $l=1,2,...., L$. Assuming that the power of transmitted signal is $P$ and  the noise at the $l$-th branch is the additive white Gaussian noise (AWGN) with zero mean and  variance $\sigma^{2}$ (assumed to be identical for all branches), then the random variable (RV) $\gamma_{l}$ is given by 
\begin{equation}
\gamma_{l}=\rho |h_{l}|^{2}, 
\label{eq_SNR}
\end{equation}
where $\rho=P/\sigma^{2}$ is the average transmit SNR, $|h_{l}|$ is a RV that represents the fading amplitude of the ${l}$-th branch and  is distributed according to a two-parameter $\mathcal{LL}$ distribution (i.e.,  $|h_{l}| \sim \mathcal{LL}(\alpha^{\prime}_{l}, \beta^{\prime}_{l})$. Hence, the
cumulative distribution function (CDF) of $|h_{l}|$ is given by \cite{9580415}
\begin{equation}
F_{|h_{l}|}(\gamma)=\frac{1}{1+ \left( \frac{\gamma}{\alpha^{\prime}_{l}}\right)^{-\beta^{\prime}_{l}} } \ \ \gamma\geq 0, 
\label{eq_channel CDF}
\end{equation}
where $\alpha^{\prime}_{l}>0$  and $\beta^{\prime}_{l}> 0$ are the scale and shape parameters of the $\mathcal{LL}$ distribution, respectively. 

Using the transformation of random variables, one can show that $\gamma_{l}  \sim \mathcal{LL}(\rho \alpha_{l}, \beta_{l})$, where $\alpha_{l}= (\alpha^{\prime}_{l})^{2}$ and  $\beta_{l}= \beta^{\prime}_{l}/2$ \footnote{ The parameters $\alpha_{l}$ and $\beta_{l}$ capture the turbulence fading characteristics of the ${l}$-th branch \cite{9580415}. The impact of these parameters on the performance of a single-branch receiver has been studied in  \cite{electronics11152409}.}. Therefore, the CDF of $\gamma_{l}$ can be obtained as 
\begin{equation}
F_{\gamma_{l}}(\gamma)=\frac{1}{1+ \left( \frac{\gamma}{\rho \alpha_{l}}\right)^{-\beta_{l}} }.
\label{eq_SNR CDF}
\end{equation}
Utilizing \cite[Eq. (8.4.2.5) and Eq. (8.2.2.14)]{prudnikovVol3}, $F_{\gamma_{l}}(\gamma)$ can be expressed as    
\begin{equation}
	\begin{split}
		F_{\gamma_{l}}(\gamma)= \Gfun*{1}{1}{1}{1}{1}{1}{ \frac{\gamma^{\beta_l}}{\left(\rho \alpha_l  \right) ^{\beta_l} }},   
	\end{split}
\label{eq_SNR CDF G}
\end{equation}
where $G_{p,q} ^{m,n}(\cdot )$ is the Meijer G-function \cite[ Eq. (8.2.1.1)] {prudnikovVol3}.

\subsection{ $L$  i.n.i.d.  branches}
We first consider an SC receiver with $L$ i.n.i.d. branches, where $\gamma_{l}$ are i.n.i.d. RVs, $l=1,2,...., L$. In view of $ \gamma_{SC}=\max\{ \gamma_{1},  \gamma_{2}, . . . . , \gamma_{L} \}$, the CDF of $ \gamma_{SC}$, denoted by  $F_{\gamma_{SC}}(\gamma)$, is given by 
\begin{equation}
			F_{\gamma_{SC}}(\gamma) = \prod_{l=1}^L F_{\gamma_l}(\gamma) = \prod_{l=1}^L \Gfun*{1}{1}{1}{1}{1}{1}{ \frac{\gamma^{\beta_l}}{\left(\rho \alpha_l  \right) ^{\beta_l} }}.
		\label{eq_CDF_SC1}
\end{equation}
$F_{\gamma_{SC}}(\gamma)$ above can be compactly expressed in terms of the multivariate H-function \cite[Definition A.1]{mathai2009h} as given in \eqref{eq_CDF_SC2}.  Capitalizing on  $F_{\gamma_{SC}}(\gamma)$ in  \eqref{eq_CDF_SC2},  we derive the probability density function (PDF) of  $ \gamma_{SC}$ in  Lemma 1 below. 
\begin{lemma}
Let $f_{\gamma_{SC}}(\gamma)$ denote the PDF of $ \gamma_{SC}=\max\{ \gamma_{1},  \gamma_{2}, . . . . , \gamma_{L} \}$, then $f_{\gamma_{SC}}(\gamma)$  is given by \eqref{eq_pdf}.\\
\textit{Proof:} See Appendix \ref{App_PDF}.  \hfill $\Box$ 
\end{lemma}

\subsection{ $L$  i.i.d.  branches}
 We now consider  an SC receiver with $L$ i..i.d branches as a special case, where $\gamma_{l}$ are i.i.d. RVs, $l=1,2,...., L$. This implies that  $\alpha_{l}=\alpha$ and $\beta_{l}=\beta$, $l=1,2,...., L$. Making use of \eqref{eq_SNR CDF} in \eqref{eq_CDF_SC1}, $F_{\gamma_{SC}}(\gamma)$ reduces to 
\begin{equation}
F_{\gamma_{SC}}(\gamma)=\frac{1}{\left[ 1+ \left( \frac{\gamma}{\rho \alpha}\right)^{-\beta} \right]^{L}} \, , \ \ \gamma\geq 0.
\label{eq_SNR CDFiid}
\end{equation}
$F_{\gamma_{SC}}(\gamma)$ can be expressed in terms of a Meijer G-function  as
\begin{equation}
	\begin{split}
F_{\gamma_{SC}}(\gamma) = \frac{1}{\Gamma(L)} \Gfun*{1}{1}{1}{1}{1}{L}{ \frac{ \gamma^{\beta}}{(\rho \alpha)^{\beta}} }, 
	\end{split}
\label{eq_SNR CDFiid_G}
\end{equation} 
where $\Gamma(\cdot)$ is the gamma function  \cite[Eq. (8.310.1)]{gradshteyn2014table}, and the identity above is obtained with the help of  \cite[Eq. (8.4.2.5) and Eq. (8.2.2.14)]{prudnikovVol3}.  To this end,  $f_{\gamma_{SC}}(\gamma)$  can be obtained by differentiating $F_{\gamma_{SC}}(\gamma)$ with respect to (w.r.t.) $\gamma$ as \cite[Eq. (8.2.2.36)]{prudnikovVol3}
\begin{equation}
	\begin{split}
f_{\gamma_{SC}}(\gamma) = \frac{\beta \gamma^{-1}}{\Gamma(L)} \Gfun*{1}{1}{1}{1}{0}{L}{ \frac{ \gamma^{\beta}}{(\rho \alpha)^{\beta}} }.
	\end{split}
\label{eq_SNR PDFiid_G}
\end{equation}

\begin{figure*}
\begin{equation}
	F_{\gamma_{SC}}(\gamma) = H_{0,0:\underbrace{1,1;\hdots;1,1}_{1,\hdots,L}}^{0,0:\overbrace{1,1;\hdots;1,1}^{1,\hdots,L}}\left[\left.\begin{smallmatrix} \left(\frac{\gamma}{\rho \alpha_1 }\right)^{\beta_1} \\ \vdots \\ \left(\frac{\gamma}{ \rho \alpha_L}\right)^{\beta_L}   \end{smallmatrix} \right| \begin{smallmatrix} - ;& { \overbrace{(1,1);\hdots;(1,1)}^{1,\hdots,L}} \\ ~&~ \\-; & \underbrace{(1,1);\hdots;(1,1)}_{1,\hdots,L} \end{smallmatrix} \right].
 \label{eq_CDF_SC2}
\end{equation}
\hrulefill
\end{figure*}

\begin{figure*}
\begin{equation}
	f_{\gamma_{SC}}(\gamma) = \frac{1}{\gamma} H_{1,1:\underbrace{1,1;\hdots;1,1}_{1,\hdots,L}}^{0,1:\overbrace{1,1;\hdots;1,1}^{1,\hdots,L}}\left[\left.\begin{smallmatrix} \left(\frac{\gamma}{\rho \alpha_1 }\right)^{\beta_1} \\ \vdots \\ \left(\frac{\gamma}{ \rho \alpha_L}\right)^{\beta_L}   \end{smallmatrix} \right| \begin{smallmatrix} (0;\beta_1,\hdots,\beta_L):& { \overbrace{(1,1);\hdots;(1,1)}^{\BraceTxt}} \\ ~&~ \\(1;\beta_1,\hdots,\beta_L): & \underbrace{(1,1);\hdots;(1,1)}_{\BraceTxt} \end{smallmatrix} \right]. 
	\label{eq_pdf}
\end{equation}
\hrulefill
\end{figure*}

\section{ Exact Performance Analysis}
\subsection{Outage Probability}
A communication outage occurs if $\gamma_{SC}$ falls below predefined threshold value $ \gamma_{th}$. Accordingly, the outage probability, denoted by $\Pout$, is defined as 
\begin{equation}
\Pout=\Pr(\gamma_{SC} \leq \gamma_{th}) =F_{\gamma_{SC}}(\gamma_{th}). 
\end{equation}
$\Pout$ can be evaluated from \eqref{eq_CDF_SC2} and \eqref{eq_SNR CDFiid} for the cases of $L$ i.n.i.d. branches and $L$ i.i.d. branches, respectively.

\subsection{Average BER}
Considering the output SNR of the SC receiver, the average BER for different  modulation formats, denoted by $\Pe$, is given by \cite{simon2005digital}
\begin{equation}
		\Pe = \int_{0}^{\infty} \delta ~\erfc\left(\sqrt{\zeta \gamma}\right) f_{\gamma_{SC}}(\gamma) d \gamma,  
	\label{eq_BER_1}
\end{equation}
where $\delta$ and $\zeta$ are modulation-dependent parameters \cite{simon2005digital} and $\erfc (\cdot)$ is complementary error function \cite[Eq. (8.25.4)]{gradshteyn2014table}.  

\begin{theorem}	
For the SC receiver with $L$ i.n.i.d. branches, the average BER is given in \eqref{eq_ABER}.\\
	\textit{Proof:} See Appendix \ref{App_BER}. \hfill $\Box$ 
\end{theorem}

\begin{corollary}
	 For the SC receiver with $L$ i.i.d. branches, the average BER in \eqref{eq_BER_1}  can be expressed as
	\begin{equation}
		\Pe = \frac{\delta}{\sqrt{\pi} \Gamma(L)} H_{4,3}^{2,3}\left[ \left. \frac{1}{\zeta \alpha \rho}  \right| \begin{smallmatrix} (0,1/ \beta),  (1,1),  (0.5,1)  (1,1) \\ ~\\  (L,1/ \beta),  (1,1), (0,1)   \end{smallmatrix} \right], 
		\label{eq_ABER_iid}
	\end{equation}
\end{corollary}
where  $H_{p,q} ^{m,n}[\cdot ]$ is the univariate H-function \cite[Eq. (1.2)]{mathai2009h}. 

\textit{Proof:} We  express the $\erfc\left(\sqrt{\zeta \gamma}\right)$ in \eqref{eq_BER_1} in terms of  Meijer G-function as  $\erfc\left(\sqrt{\zeta \gamma}\right)= \frac{1}{\sqrt{\pi}} \Gfun*{1}{1}{1}{1}{0,~1}{0,~ \frac{1}{2},~0}{\zeta \gamma }$, according to \cite[Eq. (8.4.14.2) and Eq. (8.2.2.8)]{{prudnikovVol3}}. We now replace the Meijer G-function of $\erfc\left(\sqrt{\zeta \gamma}\right)$ and $f_{\gamma_{SC}}(\gamma)$ in \eqref{eq_SNR PDFiid_G}  by their equivalent H-functions using \cite{srivastava1984treatise}. Applying \cite[Eq. (2.25.1.1)]{prudnikovVol3} to solve the integral in \eqref{eq_BER_1} yields $\Pe$ in \eqref{eq_ABER_iid}.    \hfill $\Box$

\subsection{Ergodic Capacity}
In view of $\gamma_{SC}$ being the 
output SNR of the SC receiver, the ergodic capacity, denoted by $\Ce$,  is given as 
\begin{equation}
\Ce = \frac{1}{\ln(2)}\int_0^\infty \ln(1+\gamma)f_{\gamma_{SC}}(\gamma) d\gamma.
\label{eq_Cap_Gamma}
\end{equation}

\begin{theorem}
For the SC receiver with $L$ i.n.i.d. branches, the ergodic capacity is given in \eqref{eq_AergC}.\\
\textit{Proof:} See Appendix \ref{App_ergC}. \hfill $\Box$
\end{theorem}

\begin{corollary}
For the SC receiver with $L$ i.i.d. branches, the ergodic capacity in \eqref{eq_Cap_Gamma} can be expressed as
\begin{equation}
\Ce = \frac{1}{\ln(2) \Gamma(L)} H_{3,3}^{3,2}\left[ \left. \frac{1}{ \alpha \rho} \right| \begin{smallmatrix} (0,1/ \beta), (0,1), (1,1) \\ ~\\ (L,1/ \beta), (0,1), (0,1) \end{smallmatrix} \right], 
\label{eq_AergC_iid}
\end{equation}
\end{corollary}

\textit{Proof:} We first express the $\ln(1+\gamma)$ in \eqref{eq_Cap_Gamma} in terms of Meijer G-function as $\ln(1+\gamma)= \Gfun*{1}{2}{2}{2}{1,1}{1,0}{\gamma }$, according to \cite[Eq. (8.4.6.5)]{prudnikovVol3}. To this end, we replace the Meijer G-function of $\ln(1+\gamma)$ and $f_{\gamma_{SC}}(\gamma)$ in \eqref{eq_SNR PDFiid_G} by their equivalent H-functions using \cite{srivastava1984treatise}. Applying \cite[Eq. (2.25.1.1)]{prudnikovVol3} to solve the integral in \eqref{eq_Cap_Gamma} yields $\Ce$ in \eqref{eq_AergC_iid}. \hfill $\Box$

\begin{figure*}
\begin{equation}
	\Pe = \frac{\delta}{\sqrt{\pi}} H_{1,0:\underbrace{1,1;\hdots;1,1}_{1,\hdots,L}}^{0,1:\overbrace{1,1;\hdots;1,1}^{1,\hdots,L}}\left[\left.\begin{smallmatrix} \left(\delta  \rho \alpha_1 \right)^{-\beta_1} \\ \vdots \\  \left(\delta  \rho \alpha_L \right)^{-\beta_L}  \end{smallmatrix} \right| \begin{smallmatrix} \left(\frac{1}{2};\beta_!,\hdots,\beta_L\right)& :{ \overbrace{(1,1);\hdots;(1,1)}^{1,\hdots,L}} \\ ~&~ \\- &: \underbrace{(1,1);\hdots;(1,1)}_{1,\hdots,L} \end{smallmatrix} \right].
	\label{eq_ABER}
\end{equation}
\hrulefill
\end{figure*}

\begin{figure*}[h!]
	\begin{equation}
		\Ce = \frac{1}{\ln(2)} H_{3,3:\underbrace{1,1;\hdots;1,1}_{1,\hdots,L}}^{2,2:\overbrace{1,1;\hdots;1,1}^{1,\hdots,L}}\left[\left.\begin{smallmatrix} \left(\rho \alpha_1 \right)^{-\beta_1} \\ \vdots \\  \left(\rho \alpha_L \right)^{-\beta_L}  \end{smallmatrix} \right| \begin{smallmatrix} \left(0;\beta_1,\hdots,\beta_L\right) \left(0;\beta_1,\hdots,\beta_L\right) \left(1;\beta_1,\hdots,\beta_L\right)& :{ \overbrace{(1,1);\hdots;(1,1)}^{1,\hdots,L}} \\ ~&~ \\ \left(0;\beta_1,\hdots,\beta_L\right) \left(0;\beta_1,\hdots,\beta_L\right) \left(1;\beta_1,\hdots,\beta_L\right) &: \underbrace{(1,1);\hdots;(1,1)}_{1,\hdots,L} \end{smallmatrix} \right].
		\label{eq_AergC}
	\end{equation}
	\hrulefill
\end{figure*}

\section{Asymptotic Performance Analysis}

\subsection{Asymptotic Outage probability}
To get further insights, we derive next an asymptotic closed-form expression for $F_{\gamma_l}(\gamma)$ in \eqref{eq_SNR CDF} at high values of average transmit SNR (i.e., $\rho \to \infty$). It can be shown that 
\begin{equation}
F_{\gamma_l}(\gamma) \approx \frac{1}{\left(\rho \alpha_l \right)^{\beta_l}} \gamma^{\beta_l},
\label{eq_CDFasySB}
\end{equation}
as $\rho \to \infty$. 
Accordingly, the asymptotic CDF of $\gamma_{SC}$ for the case of $L$ i.n.i.d. branches can be given as 
\begin{equation}
F_{\gamma_{SC}}(\gamma) = \prod_{l=1}^L F_{\gamma_l}(\gamma) \approx \varphi \rho^{-\Sb}\gamma^{\Sb},
\label{eq_CDFasy}
\end{equation}
where $\Sb = \sum_{l=1}^L \beta_l$ and $\varphi = \prod_{l=1}^L \left(\frac{1}{\alpha_l}\right)^{\beta_l}$. Consequently, the asymptotic PDF of $\gamma_{SC}$ is 
\begin{equation}
f_{\gamma_{SC}}(\gamma) \approx \varphi \Sb \ \rho^{-\Sb} \gamma^{\Sb -1}. 
\label{eq_PDFasy}
\end{equation}

Capitalizing on \eqref{eq_CDFasy}, the asymptotic outage probability, denoted by $\Poutasy$, can be obtained as 
\begin{equation}
\Poutasy \approx \varphi \rho^{-\Sb} \gamma_{th}^{\Sb}. 
\label{eq_Poutasy}
\end{equation}

\subsection{Asymptotic Average BER }
The asymptotic average BER, denoted by $\Peasy$, can be obtained by plugging $f_{\gamma_{SC}}$ in \eqref{eq_PDFasy} and the identity $\erfc\left(\sqrt{\zeta \gamma}\right)= \frac{1}{\sqrt{\pi}} \Gfun*{1}{1}{1}{1}{0,~1}{0,~ \frac{1}{2},~0}{\zeta \gamma }$ in the integral representation of \eqref{eq_BER_1}. To this end, the integral in \eqref{eq_BER_1} can be solved with the help of \cite[Eq. (2.24.2.1)]{prudnikovVol3} as
\begin{equation}
\Peasy \approx \frac{\delta\zeta^{-\Sb}}{\sqrt{\pi}} \varphi \Gamma\left(\frac{1}{2}+\Sb\right) \rho^{-\Sb}. 
\label{eq_Peasy1}
\end{equation} 

Based on \eqref{eq_Poutasy} and \eqref{eq_Peasy1}, we can draw conclusions about the diversity order $G_{d}$ ( see, e.g., \cite{electronics11152409} and references therein). It can be shown that for the  i.n.i.d. case $G_{d}=S_{\beta}=\sum_{l=1}^{L}\beta_{l}$, while for the i.i.d. case $G_{d}=S_{\beta}= \beta L$.

\subsection{Asymptotic Capacity}
The asymptotic ergodic capacity, denoted by $\Ceasy$, can be evaluated according to \cite{yilmaz2012novel} as
\vspace{-1mm}
\begin{equation}\label{eq_C1_asy}
\Ceasy \approx \log _{2}\left({\rho}\right)+\left.\log _{2}(e) \frac{\partial}{\partial n} \frac{\mathbb{E}\left[\gamma^{n}_{SC}\right]}{{\rho}^{n}}\right|_{n=0},
\end{equation}
where the $\mathbb{E}\left[\gamma^{n}_{SC}\right]$ is the $n$-th moment of the RV $\gamma_{SC}$. To this end, we utilize $F_{\gamma_{SC}}(\gamma)$ in \eqref{eq_SNR CDFiid} and \cite[Eq. (3.241.4)]{gradshteyn2014table} to evaluate the $n$-th moment of $\gamma_{SC}$ with $L$ i.i.d. branches as
\begin{equation}
\begin{split}
\mathbb{E}\left[\gamma_{SC}^{n}\right] = {\frac {{\rho}^{n}{\alpha}^{n}}{\Gamma \left( L \right) }\Gamma \left( {\frac {\beta-n}{\beta}} \right) \Gamma \left( {\frac {\beta\, L+n}{\beta}} \right) }.
\end{split}
\label{eq_En}
\end{equation}

Plugging \eqref{eq_En} in \eqref{eq_C1_asy}, after some basic algebraic manipulations, yields
\begin{equation}
\Ceasy \approx \log _{2}\left({\rho}\right)+ \frac{\beta \ln(\alpha)+E_{0}+\Psi(L)}{\beta \ln(2)}, 
\end{equation}
where $E_{0}$ denotes the Euler-Mascheroni constant and $\Psi(\cdot )$ is the digamma function \cite[Eq. (8.360.1)]{gradshteyn2014table}.

\section{Application Example}
In this section, we consider an UWOC system affected by temperature-induced turbulence \cite{9580415}. We utilize the derived expressions for the outage probability, the average BER, and  the ergodic capacity to investigate numerically the performance of such a system. To verify our analytical results for different system configurations, we consider the following  scenarios: 
\begin{itemize}
  \item Scenario 1)  $L$ i.n.i.d. branches SC receiver is employed when $L=3$. The channel power gain of each branch is distributed according to a $\mathcal{LL}$ distribution with arbitrarily chosen parameters $(\alpha_{l}, \beta_{l}) \in \{ (1, 2.2), (0.98, 2.3), (1.1, 2.4) \}$. 

  \item Scenario 2) $L$  i.i.d. branches SC receiver is employed when $L=2$ and $L=4$. The channel power gain of each branch is distributed according to a $\mathcal{LL}$ distribution with measurement-based parameters $\alpha=0.9724, \beta=2.3311$, as given in Case 2 of Table II \cite{9580415}.  
\end{itemize}

In Fig. 1, we plot the outage probability against the average transmit SNR $\rho$ for different values of $L$  when $\gamma_{th}=10$ dB. Fig. 2 shows the average BER versus $\rho$, where the intensity modulation and direct detection (IM-DD) with on-off keying (OOK) modulation is employed. Fig. 3 depicts the ergodic capacity versus $\rho$ for different values of $L$. In all figures, we observe that the analytical results perfectly match with the simulation results and the asymptotic results tend to converge to the exact ones as  $\rho$ grows large. In addition, increasing $L$ improves the performance of the system, as expected.

\begin{figure*}[!t]
    \centering
\begin{minipage}[t]{0.32\linewidth} 
\centering
    \includegraphics[width=1\linewidth]{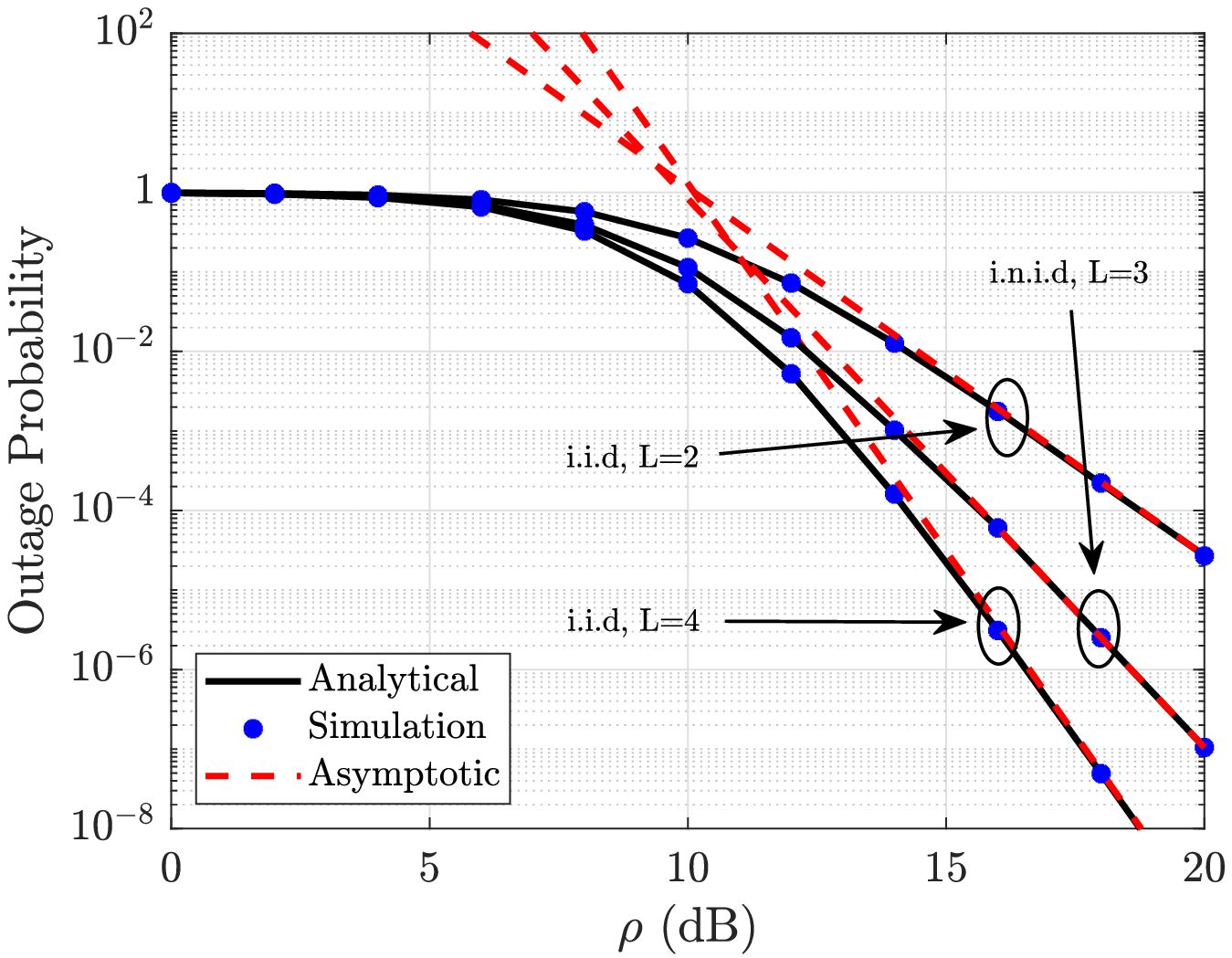}
    \caption{Outage Probability versus $\rho$. }
    \label{fig:img1}
\end {minipage}
\begin{minipage}[t]{0.32\linewidth} 
\centering
    \includegraphics[width=1\linewidth]{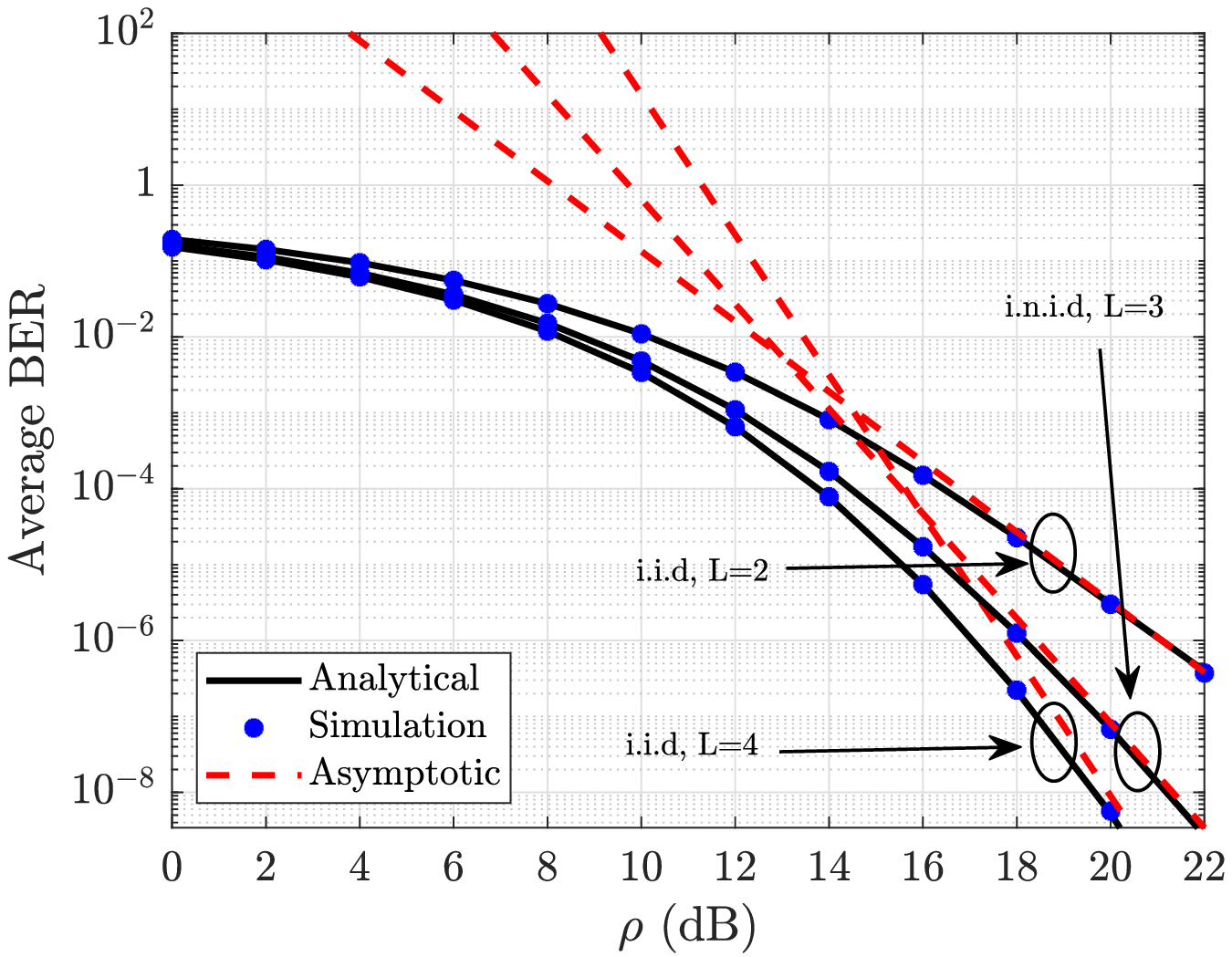}
    \caption{Average BER versus $\rho$.}
    \label{fig:img1}
\end {minipage}
\begin{minipage}[t]{0.32\linewidth} 
\centering
    \includegraphics[width=1\linewidth]{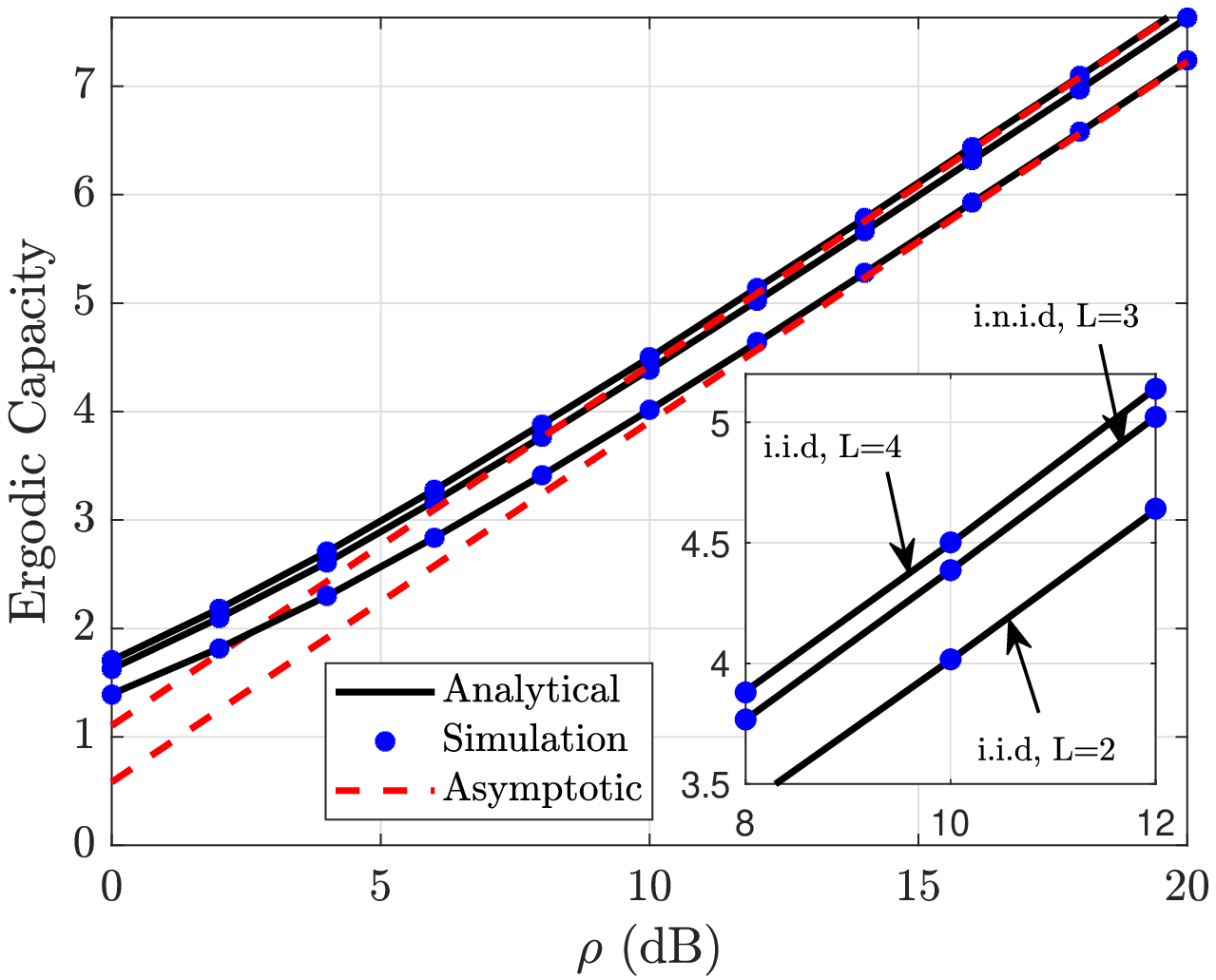}
    \caption{Ergodic Capacity versus $\rho$.}
    \label{fig:img1}
\end {minipage}
\end{figure*}

\section{Conclusion}  
The performance of an SC receiver operating over i.n.i.d. log-logistic fading channels is investigated. We developed exact analytical expressions for the outage probability, the average BER, and the ergodic capacity. We also obtained exact expressions for these measures for i.i.d. channels, as a special case. Furthermore, we obtained simplified asymptotic expressions for such performance measures in the high SNR regime. We utilized  the analytical results to analyze the performance of an SC receiver in UWOC systems. Monte Carlo Simulations have confirmed the correctness of our analytical framework.

\appendices
\section{Derivation  of $f_{\gamma_{SC}}(\gamma)$} \label{App_PDF}
Re-expressing the H-function of $F_{\gamma_{SC}}(\gamma)$ in \eqref{eq_CDF_SC2} by its original contour integral according to \cite[Definition A.1]{mathai2009h}, yields 
\begin{equation}
	\begin{split}
		F_{\gamma_{SC}}(\gamma)   = \left( \frac{1}{2\pi j}\right)^L &\oint\limits_{c_1} \hdots \oint\limits_{c_L}   \left[ \Pi_{l=1}^L \Gamma(1-s_i) \Gamma(s_i)  \right. \\ &\left.(\rho \alpha_l )^{-\beta_l}  \right]   \gamma^{\sum_{l=1}^L \beta_l} ds_1 \hdots ds_L. 
	\end{split}
\end{equation}	
Then, $f_{\gamma_{SC}}(\gamma)$ is obtained by differentiating $F_{\gamma_{SC}}(\gamma)$ w.r.t. $\gamma$ as 
\begin{equation}
	\begin{split}
		f_{\gamma_{SC}}(\gamma) &= \frac{\partial F_{\gamma_{SC}}(\gamma) }{\partial \gamma}  = \left( \frac{1}{2\pi j}\right)^L \oint\limits_{c_1} \hdots \oint\limits_{c_L}   \left[ \Pi_{l=1}^L \Gamma(1-s_i)    \right. \\ &  \left. \Gamma(s_i) ( \rho \alpha_l)^{-\beta_l s_l}  \right]   \left(\frac{\partial}{\partial \gamma}  \gamma^{\sum_{l=1}^L \beta_l s_l}  \right) ds_1 \hdots ds_L. 
	\end{split}
	\label{eq_pdf_1}
\end{equation}
To this end, 
\begin{equation}
	\begin{split}
		\frac{\partial}{\partial \gamma}  \gamma^{\sum_{l=1}^L \beta_l s_l} & =    \left(\sum_{l=1}^L \beta_l s_l \right) \gamma^{\left(\sum_{l=1}^L \beta_l s_l \right) - 1} \\ &= \frac{\Gamma(1+\sum_{l=1}^L \beta_l s_l)}{\Gamma(\sum_{l=1}^L \beta_l s_l)} \gamma^{\left(\sum_{l=1}^L \beta_l s_l\right) - 1}, 
	\end{split} 
\label{eq_der_gamma}
\end{equation}
where the second equality in \eqref{eq_der_gamma} is obtained with the help of $\Gamma(1+x) = x \Gamma(x)$  \cite[Eq. (8.331.1)]{gradshteyn2014table} with $x = \sum_{l=1}^L \beta_l s_l$. Plugging \eqref{eq_der_gamma} in \eqref{eq_pdf_1} yields \eqref{eq_pdf_contour}. Utilizing \eqref{eq_pdf_contour} and \cite[Definition A.1]{mathai2009h}, $f_{\gamma_{SC}}(\gamma)$ is as given in \eqref{eq_pdf}.
\begin{figure*}
	\hrulefill
	\begin{equation}
	\begin{split}
		f_{\gamma_{SC}}(\gamma) &= \left( \frac{1}{2\pi j}\right)^L \oint\limits_{c_1} \hdots \oint\limits_{c_L}      \frac{\Gamma(1+\sum_{l=1}^L \beta_l s_l)}{\Gamma(\sum_{l=1}^L \beta_l s_l)} \gamma^{\left(\sum_{l=1}^L \beta_l s_l\right) - 1} \left[ \prod_{l=1}^L \Gamma(1-s_i) \Gamma(s_i) (\rho \alpha_l )^{-\beta_l s_l}  \right] ds_1 \hdots ds_L \\ &= \frac{1}{\gamma} \left( \frac{1}{2\pi j}\right)^L \oint\limits_{c_1} \hdots \oint\limits_{c_L}   \frac{\Gamma(1+\sum_{l=1}^L \beta_l s_l)}{\Gamma(\sum_{l=1}^L \beta_l s_l)}  \left[ \prod_{l=1}^L \Gamma(1-s_i) \Gamma(s_i) \left(\frac{\gamma}{\rho \alpha_l }\right)^{\beta_l s_l}  \right]    ds_1 \hdots ds_L.
	\end{split}
	\label{eq_pdf_contour}
\end{equation}
	\hrulefill
\end{figure*}

\section{Derivation of the BER} \label{App_BER}
Plugging  \eqref{eq_pdf_contour} in \eqref{eq_BER_1} yields \eqref{eq_BER_2}. If we Let $z = \sqrt{\zeta \gamma}$, then the inner integral $I_1$ in \eqref{eq_BER_2} can be written as
\begin{figure*}
\vspace{-6mm}
\begin{equation}
	\begin{split}
		\Pe  & = \left( \frac{1}{2\pi j}\right)^L \oint\limits_{c_1} \hdots \oint\limits_{c_L} \left[ \Pi_{l=1}^L \Gamma(1-s_l) \Gamma(s_l) (\rho \alpha_l )^{-\beta_l s_l}  \right] \frac{\Gamma(1+\sum_{l=1}^L \beta_l s_l)}{\Gamma(\sum_{l=1}^L \beta_l s_l)} \\& \hspace{3cm}  \times \underbrace{\left(\int_{0}^{\infty} \delta \gamma^{\left(\sum_{l=1}^L \beta_l s_l \right) - 1} ~\erfc\left(\sqrt{\zeta \gamma}\right)     d\gamma \right) }_{I_1} ds_1 \hdots ds_L. 
	\end{split}
\label{eq_BER_2}
\end{equation}
	\hrulefill
\end{figure*}

\begin{figure*}[t!]
\vspace{-6mm}
	\begin{equation}
		\begin{split}
			C  & = \frac{1}{\ln(2)} \left( \frac{1}{2\pi j}\right)^L \oint\limits_{c_1} \hdots \oint\limits_{c_L} \left[ \Pi_{l=1}^L \Gamma(1-s_l) \Gamma(s_l) (\rho \alpha_l )^{-\beta_l s_l}  \right] \frac{\Gamma(1+\sum_{l=1}^L \beta_l s_l)}{\Gamma(\sum_{l=1}^L \beta_l s_l)} \\& \hspace{3cm}  \times \underbrace{\left(\int_{0}^{\infty}  \gamma^{\left(\sum_{l=1}^L \beta_l s_l \right) - 1} ~\ln\left(  1+ \gamma \right)     d\gamma \right) }_{I_2} ds_1 \hdots ds_L. 
		\end{split}
		\label{eq_C_2}
	\end{equation}
	\hrulefill
\end{figure*}

\begin{figure*}[t!]
\vspace{-4mm}
	\begin{equation}
		\begin{split}
			C  & = \frac{1}{\ln(2)}\left( \frac{1}{2\pi j}\right)^L \oint\limits_{c_1} \hdots \oint\limits_{c_L} \frac{\Gamma(1+\sum_{l=1}^L \beta_l s_l)\Gamma(-\sum_{l=1}^L \beta_l s_l)\Gamma(1+\sum_{l=1}^L \beta_l s_l)\Gamma(-\sum_{l=1}^L \beta_l s_l)}{\Gamma(\sum_{l=1}^L \beta_l s_l)\Gamma(1-\sum_{l=1}^L \beta_l s_l)} \\& \hspace{3cm}  \times  \left[ \Pi_{l=1}^L \Gamma(1-s_l) \Gamma(s_l) (\rho \alpha_l )^{-\beta_l s_l}  \right] ds_1 \hdots ds_L. 
		\end{split}
		\label{eq_C_3}
	\end{equation}
	\hrulefill
\end{figure*}
\begin{equation}
	I_{1} = \frac{2\delta}{\zeta^{\sum_{l=1}^L \beta_l s_l}} \int_{0}^{\infty}  z^{\left(2\sum_{l=1}^L \beta_l s_l \right) - 1} ~\erfc\left(z\right)  dz. 
\end{equation}  

Utilizing \cite[Eq. (2.8.2.1)]{prudnikov1986special}, $I_{1}$ above can be expressed as
\begin{equation}
	\begin{split}
		I_1 &= \frac{\delta ~ \Gamma\left( \frac{1}{2} + \sum_{l=1}^L \beta_l s_l \right) }{\sqrt{\pi}  \zeta^{\sum_{l=1}^L \beta_l s_l} \left(\sum_{l=1}^L \beta_l s_l \right)  }  \\ &=\frac{\delta }{\sqrt{\pi}  \zeta^{\sum_{l=1}^L \beta_l s_l}  }  \frac{\Gamma\left( \frac{1}{2} + \sum_{l=1}^L \beta_l s_l \right) \Gamma\left( \sum_{l=1}^L \beta_l s_l \right)}{\Gamma\left( 1 + \sum_{l=1}^L \beta_l s_l \right) },
	\end{split}
\label{eq_I}
\end{equation}
where the second equality is obtained with the help of $\Gamma(1+x) = x \Gamma(x)$  with $x = \sum_{l=1}^L \beta_l s_l$. Plugging \eqref{eq_I} in \eqref{eq_BER_2} yields
\begin{equation}
	\begin{split}
		\Pe  & = \frac{\delta }{\sqrt{\pi}} \left( \frac{1}{2\pi j}\right)^L \oint\limits_{c_1} \hdots \oint\limits_{c_L} \Gamma\left( \frac{1}{2} + \sum_{l=1}^L \beta_l s_l \right) \\ & \times \left[ \prod_{l=1}^L \Gamma(1-s_l) \Gamma(s_l) (\zeta \rho \alpha_l )^{-\beta_l s_l}  \right]   ds_1 \hdots ds_L. 
	\end{split}
	\label{eq_BER_contour}
\end{equation}
Utilizing \eqref{eq_BER_contour} and \cite[Definition A.1]{mathai2009h}, $\Pe$ is as in \eqref{eq_ABER}. 

\section{Derivation  of the ergodic Capacity} \label{App_ergC}
Plugging \eqref{eq_pdf_contour} in \eqref{eq_Cap_Gamma} yields the contour integral representation of the ergodic capacity in \eqref{eq_C_2}.  With the help of \cite[Eq. (8.4.6.5)]{prudnikovVol3}, the inner integral $I_2$ in \eqref{eq_C_2} can be written as
\begin{equation}
	I_2 = \int_{0}^{\infty}  \gamma^{\left(\sum_{l=1}^L \beta_l s_l \right) - 1} \Gfun*{1}{2}{2}{2}{1,1}{1,0}{\gamma }     d\gamma. \\
\end{equation}
Utilizing  \cite[Eq. (2.24.2.1)]{prudnikovVol3}, $I_2$ can be expressed as
\begin{equation}
	I_2	= \frac{\Gamma\left(1 + \sum_{l=1}^L \beta_l s_l \right) \Gamma\left( -\sum_{l=1}^L \beta_l s_l \right) \Gamma\left( -\sum_{l=1}^L \beta_l s_l \right)}{\Gamma\left( 1 - \sum_{l=1}^L \beta_l s_l \right) }.
	\label{eq_I2}
\end{equation} 
Substituting \eqref{eq_I2} in \eqref{eq_C_2} leads to \eqref{eq_C_3}.  Utilizing  \eqref{eq_C_3} and \cite[Definition A.1]{mathai2009h},  the ergodic capacity is as in \eqref{eq_AergC}. 
\bibliographystyle{IEEEtran}
\bibliography{Albadarneh}

\end{document}